 \documentclass[preprint]{aastex}

\newcommand{\vvm}{$V/V_{\rm max}$ }
\newcommand{\avvvm}{$<V/V_{\rm max}>$ }
\newcommand{\alp}{$\alpha_{23}$ }
\newcommand{\lfsp}{$\Phi_0(L,sp)$ }
\newcommand{\lf}{$\Phi_0(L)$ }
\newcommand{\sfr}{$R_{\rm GRB}(z)$ }
\newcommand{\np}{$\log N-\log P$ }

\shorttitle{Schmidt}
\shortauthors{Luminosity Function of Gamma-Ray Bursts}

\begin{document}

\title{Luminosity Function of Gamma-Ray Bursts \\
       Derived Without Benefit of Redshifts}

\author{Maarten Schmidt}
\affil{California Institute of Technology, Pasadena, CA 91125}
\email{mxs@deimos.caltech.edu}

\begin{abstract}
We show that the Euclidean value of \avvvm for gamma-ray bursts (GRB) 
selected on a timescale of 1024 ms is correlated with spectral hardness. 
The value of \avvvm ranges from $\sim 0.42$ for soft bursts to 
$\sim 0.26$ for the hardest bursts. Given that the Euclidean value 
of \avvvm for cosmological objects in a well defined sample is a 
distance indicator, the hard bursts must reside at larger redshifts
and therefore be more luminous than the soft bursts. The resulting
luminosity-hardness correlation cannot be shown explicitly due to 
the small number of observed GRB redshifts at the present time. 
Based on the $<V/V_{\rm max}>$-hardness
correlation, we derive the luminosity function of GRBs without 
using any redshifts, but we have to make an assumption how the 
comoving GRB rate varies with redshift. We present luminosity functions 
for three models of the GRB rate as a function of redshift, based on 
star formation rates. The peak luminosity functions 
are approximately broken power laws with an isotropic-equivalent break 
luminosity of $\sim 10^{51.5}$ erg s$^{-1}$ in the $50-300$ keV range 
and total local rate densities of $\sim 0.5$ Gpc$^{-3}$ y$^{-1}$.
Predicted GRB counts as a function of flux and redshift are presented.
Based on the GRB luminosity function, we carry out a simulation to 
produce the luminosity-hardness correlation, which shows that the hardest 
GRBs are $\sim 20$ times more luminous than the softest ones.

\end{abstract}


\section{Introduction}

Since there are only a small number of gamma-ray bursts (GRBs) with
observed redshifts at the present time, the derivation of their
luminosity function has presented a difficult problem. A well 
established GRB luminosity function would allow predictions of the 
distributions of flux and of redshift for planned GRB surveys. 
The space densities and luminosities of GRBs would be 
useful in discussing their relation to other objects and
in understanding the physics of these fascinating objects. Usually, the
luminosity function of extragalactic objects is derived from a set of
samples of the objects with observed redshifts to well defined flux
limits. Within a given cosmological model, such observations allow a
derivation of the luminosity function, including its dependence on
redshift. 

The Burst and Transient Source Experiment (BATSE) 
\citep{fis89} on board the {\it Compton Gamma-Ray Observatory} has
made possible the collection of large well defined GRB samples, such
as the BATSE catalog\footnote{The BATSE catalog (C. A. Meegan et al.
1999) is available at http://gammaray.msfc.nasa.gov/batse/grb/
catalog/current/.} 
based on on-board triggers, and the BD2 sample \citep{sch99a,sch99b} 
based on BATSE DISCLA data (see Sec. 2).
Redshifts are available for a relatively small number of GRBs.
At the present time no well defined sample of GRBs can be formed in which 
all or a substantial fraction of the objects has an observed redshift.
As a consequence various other methods have been used to derive the GRB
luminosity function.

The observed \np relation in a survey carries information about the 
luminosity of the survey objects. For a given cosmological model,
the flux at which the \np relation 
starts to fall below the Euclidean slope of $-3/2$ allows derivation of
the corresponding luminosity. Generally, the analyses
assumed that GRBs were standard candles \citep{coh95,pen96,wij98,tot99}
and derived the maximum redshift or the standard candle luminosity by
fitting the observed \np distribution for a given cosmological
model and an assumed GRB density distribution.

A different method to derive the luminosity function makes use of a 
luminosity indicator, i.e., an observable property that correlates with 
luminosity. Several luminosity indicators have been proposed, such as 
the spectral lag derived from cross-correlation of two spectral channels
\citep{nor00}, and the variability in the time profile \citep{fen00}.
The correlations of these properties with luminosity have been
discovered and calibrated from a relatively small number of GRBs 
with redshifts, e.g., the variability correlation is based on 
seven redshifts \citep{fen00}.

We have previously explored a reverse approach, in which we made assumptions
about the shape and extent of the luminosity function, and then used 
the Euclidean \avvvm value of the BD2 sample 
(see Sec. 2) to derive the characteristic 
luminosity $L^*$ and the local GRB space density, again for a given 
cosmological model and GRB space distribution \citep{sch99b}.
Varying the assumptions about the shape of the luminosity function 
allowed an evaluation of the sensitivity of $L^*$ and the space density 
to the various input parameters.

\begin{figure}
\plotone{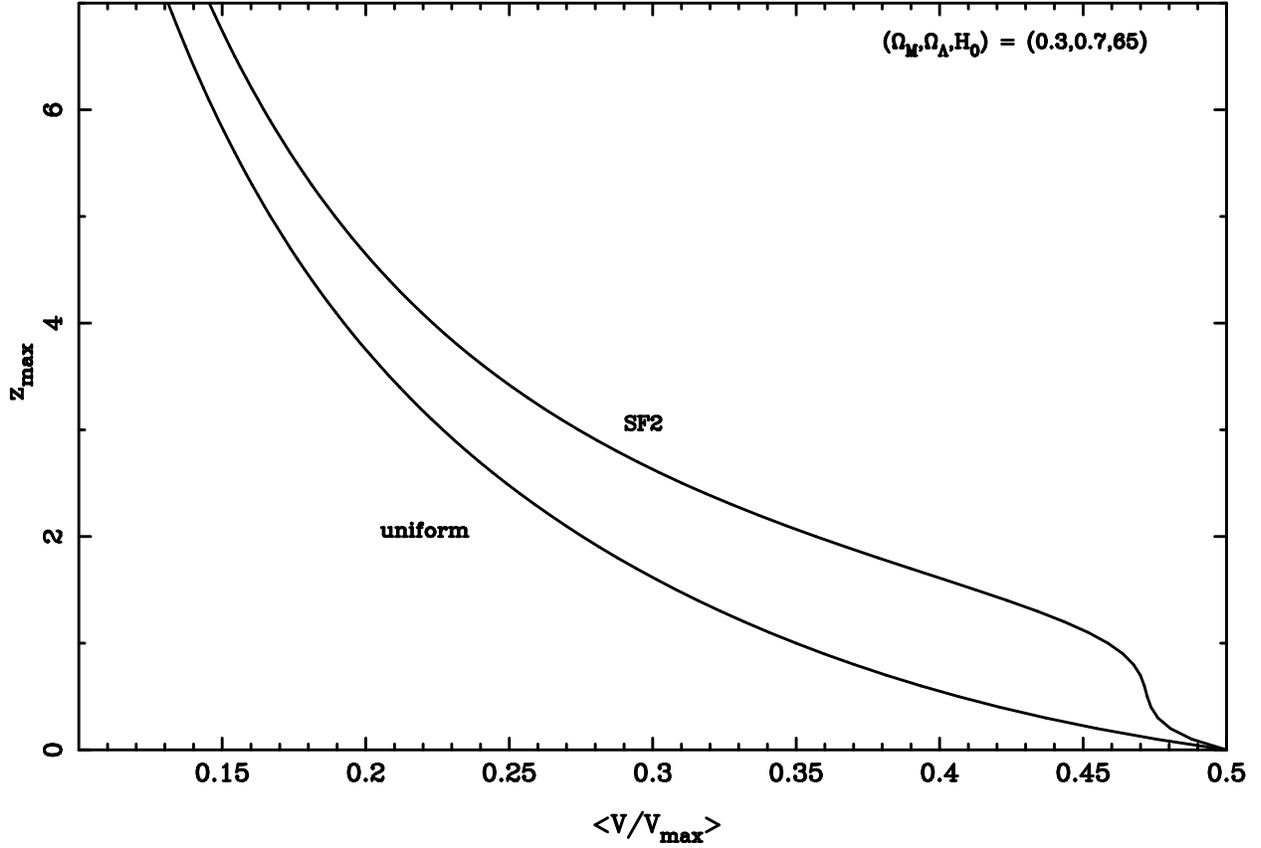}
\caption{Illustration of the use of the Euclidean value
of \avvvm as a cosmological distance indicator. The objects used are
standard candle bursts with a photon spectrum $E^{-2}$ in a cosmological
model as indicated. The co-moving space density is taken to be
uniform, or to be proportional to the star formation rate derived
by \citet{ste99} (model SF2, see Sec. 3). 
Plotted is the maximum redshift $z_{max}$ of a sample of
bursts versus the sample \avvvm. \label{fig1}}
\end{figure}

\placefigure{fig1}

The deviation of the Euclidean \avvvm of cosmological objects in a
well defined sample from the value 0.5 reflects to first order
the effect of using Euclidean geometry in its derivation rather than an
appropriate relativistic cosmological model. These effects increase 
with distance and therefore the Euclidean \avvvm is  
a distance indicator for the sample objects. We illustrate this in
Figure~\ref{fig1}, where we plot the maximum redshift of a sample of
bursts versus the Euclidean \avvvm for standard candle GRBs with 
a photon spectrum $\sim E^{-2}$. The curves show the
relation for two different scenarios for the comoving space
distribution. Clearly, the Euclidean value of \avvvm is a cosmological 
distance indicator that can be used to determine a distance scale for
cosmological objects for which no redshifts are available.

In this paper, we will demonstrate a correlation between 
the Euclidean value of \avvvm and the spectral hardness of GRBs. 
We interpret this correlation in terms of a 
luminosity-hardness correlation but are initally unable to show
the correlation explicitly. We use the \avvvm - hardness
correlation to derive the GRB luminosity function, as well as
predicted counts as a function of flux and redshift. Finally,
we show the results of a simulation producing the
luminosity-hardness correlation. 

The paper is organized as follows. In Sec. 2, we
present the observed correlation between spectral hardness and \avvvm. 
The methodology used to derive the luminosity function is discussed in
Sec. 3. The resulting luminosity functions are shown in Sec. 4, together
with predicted distributions of flux and redshift. We show the results 
of a simulation producing the luminosity-hardness correlation in
Sec. 5, followed by the discussion in Sec. 6.
Throughout this paper, we will be using a flat cosmological model with
$H_0 = 65~$km s$^{-1}$ Mpc$^{-1}$, $\Omega_M = 0.3$, 
and $\Omega_{\Lambda} = 0.7$ \citep{bah99}. 

\section{Spectral Hardness as a Luminosity Indicator}

\subsection{Data}

In this paper we use a large homogeneous sample, the BD2 sample,
derived from BATSE DISCLA data consisting of the continuous
data stream from the eight BATSE LAD detectors in four energy
channels on a timescale of 1024 ms \citep{fis89}. The sample was
derived using a software trigger algorithm that interpolated the
background between given times before and after the onset of the 
burst and required an excess of at least 5$\sigma$ over background
in at least two detectors in the energy range $50-300$ keV
\citep{sch99a}. The first version (the BD1 sample) was described
in \citet{sch99a}. A revision discussed in Sec. 2 of \citet{sch99b}
produced the BD2 sample.

The BD2 sample covers a period of 5.9 y from TJD $8365-10528$.
It contains 1391 GRBs, of which 1013 are also listed in the BATSE
catalog. The median photon flux limit of the BD2 sample over the 
energy range $50-300$ keV is 0.31 ph cm$^{-2}$ s$^{-1}$. 
The average Euclidean \vvm is $0.336\pm0.008$. The sample of 1391 
GRBs effectively represents 2.003 y of full sky coverage. 

\placefigure{fig2}

In studying the correlation of spectral hardness of GRBs with other
properties, we have to choose a relevant part of the light curve since
the spectral hardness of GRBs generally varies while the burst is going on.
The Euclidean values of \vvm in the BD2 sample have been derived from
simulations in which the GRB is moved out until the detection algorithm
fails to trigger \citep{sch99a}. In most cases, the final detection is
on the peak of the GRB time profile. Therefore, we use the 1024 ms
interval containing the peak flux to derive the hardness ratio HR32
as the ratio of the burst counts in BATSE channels 3 ($100-300$ keV)
and 2 ($50-100$ keV) for the brightest illuminated detector. From HR32, 
the illumination angle and the BATSE detector response matrix, we then
derive the photon spectrum slope \alp. 

\placetable{tbl-1}

In Figure~\ref{fig2}, we plot the Euclidean value of \vvm versus \alp
for all 1391 GRBs in the BD2 sample. We also show \avvvm for 4 spectral
classes of $\sim 348$ GRBs each (see Table~\ref{tbl-1}).
The \avvvm values range from $\sim 0.47$ for the softest bursts 
to $\sim 0.27$ for the hard ones. The mean errors of the \avvvm
values are around $\pm 0.016$.

Before we use this correlation between \avvvm and $<$\alp$>$, we
evaluate the effect of statistical errors in the counts used
to derive \alp. The
detection of GRBs in the BD2 sample was based on the sum of the
counts in channels 2 and 3, without regard to their ratio HR32.
The statistical error in HR32, determined by the burst counts and
the background counts in each of the two channels, will be larger for
weaker GRBs, which have larger values of \vvm. We see in Figure~\ref{fig2}
a strong concentration of \alp around $-1.5$ at low \vvm where 
errors are small. As we move upward in Figure~\ref{fig2}, the
statistical errors will scatter \alp horizontally, causing the
concentration at \alp$ \sim -1.5$ to widen. There will be a net
movement of points to the left and to the right and this effect 
will increase toward the top of the diagram. This causes \avvvm 
to be too large for \alp values well below and above $-1.5$. 

We have carried out simulations to estimate the systematic effect
on \avvvm. We used actually measured
background counts in channels 2 and 3 for each of the 1391 GRBs
and took the errors to be gaussian.
We assumed that the actual distribution of \alp is represented
by the brightest quartile of 348 GRBs, which are confined
to \vvm$ < 0.069$ at the very bottom of Figure~\ref{fig2}.
We ensured that in the input sample
there was no correlation between \avvvm and \alp
by associating each \vvm with each \alp, so that the input
set of bright bursts consisted of $348 \times 348$ objects.

This hypothetical bright set was then used to simulate \vvm and
\alp values for the 1043 objects with \vvm$ > 0.069$. The results 
are exhibited in Table~\ref{tbl-1} and Figure~\ref{fig2}, where we
show both observed values of $<$\alp$>$ and \avvvm, and values
corrected for the systematic errors resulting from the simulations.
For the softest and the 
hardest classes, the errors caused \alp to move away from the
center and \avvvm to increase. For the classes in the middle,
\avvvm decreased because high values of \vvm scattered away
in \alp. We are not showing mean errors for the corrected values
of \avvvm. They will be at least $\pm 0.016$ but are likely
dominated by systematic errors in the simulations or in the
statistical behavior of the observed counts that we cannot judge.

\subsection{Interpretation as Luminosity-Hardness Correlation}

Given that the Euclidean value of \avvvm is a distance indicator
(see Fig.~\ref{fig1}), the hard bursts in the sample on the average 
must be at larger distances and the softer bursts nearer. With
the density distribution SF2 (see Sec. 3) and the assumption of
standard candles used for the illustration in Figure~\ref{fig1}, 
one would estimate a maximum redshift of $\sim 1.4$ for 
the softest bursts and $\sim 3.4$ for the hardest bursts. 
With the given flux limit of the BD2 sample, the soft bursts are
therefore of lower luminosity and the hard bursts of higher luminosity. 
We conclude that the spectral index \alp is a luminosity indicator
and that there exists a luminosity-hardness correlation. 

At this stage we cannot show the luminosity-hardness correlation
explicitly due to the dearth of observed redshifts. Once we have
derived the GRB luminosity function, we will produce an explicit 
luminosity-hardness correlation through simulation, see Sec. 5.

Could the correlation of \avvvm on \alp be a consequence of the shape
of the spectrum of
GRBs? The typical GRB photon spectrum is often characterized as the Band
spectrum \citep{ban93}, with low energy slope $\sim-1$, high energy slope
$\sim-2$ and break energy near 150 keV. At low redshift, this spectrum
will have an observed \alp between $-1$ and $-2$, while at high redshift 
it will be $-2$. Thus more distant bursts will have softer spectra,
which is the opposite of our finding above.

Previous studies of the correlation between hardness 
and global GRB properties have been based on the \np relation
or on \vvm, often involving duration-hardness classes
\citep{bel92,bel96,kou93,kou96,pen98,tav98}. Among the three
classes considered by \citet{tav98}, classes B and C contained
GRBs with $T_{90} > 2.5$ s similar to the BD2 GRBs
detected at a time scale of 1024 ms. The hard bursts in class B
had \avvvm $= 0.29$ while the soft bursts in class C had
\avvvm $= 0.42$, showing the same trend as seen in Figure~\ref{fig2}.
Also of interest is the study by \citet{pen98} who defined
NHE bursts as those that have a marked lack of high-energy flux 
($E > 300$ keV), in contrast to HE bursts that have a strong 
high-energy flux. Using the \np relations for both types they concluded 
that HE bursts are eight times more luminous than NHE bursts. This
agrees qualitatively with our finding that hard bursts are more
luminous than soft bursts. 

We have not used the durations or the duration-hardness classification
of GRBs for the following reason. In the BD2 sample we define the 
duration of a GRB as the total time elapsed between the time of
trigger and the last time the burst flux exceeded the limiting flux
for triggering. In our simulations in which we move a burst out in
distance to derive its \vvm, we find that the duration decreases 
to around 1 or 2 seconds when last detected. Clearly, 
our definition of 'duration' does not produce an absolute
property of the burst. Therefore, we have not considered 
duration-hardness classes for the BD2 GRBs. The BATSE $T_{90}$ and 
$T_{50}$ durations also suffer from a ``fluence-duration'' bias 
according to \citet{hak00}. If the derived duration of a given burst
depends on its flux, then the \vvm values 
for bursts with a minimum observed duration have to be derived
using two simultaneous limits \citep{sch68}, i.e. the flux limit 
and the duration limit. Ignoring this requirement will give rise to
systematic errors in \vvm.

\section{Deriving the GRB Luminosity Function From \avvvm}

In the next section, we will find that the central luminosities of the 
4 spectral classes (see Table 1)
range over a factor of around $10-50$. This
makes it possible to construct the luminosity function of GRBs.
The derivation for each of the spectral classes is similar to that 
employed previously in \citet{sch99b} for the entire luminosity function.
We use the cosmological parameters $H_0 = 65~$km s$^{-1}$ Mpc$^{-1}$,
$\Omega_M = 0.3$, and $\Omega_{\Lambda} = 0.7$. 

We assume that the GRB luminosity function $\Phi(L,z,sp)$ 
of the spectral class $sp$ can be written as
\begin{equation}
 \Phi(L,z,sp) = \Phi_0(L,sp)R_{\rm GRB}(z),
\end{equation}
where $sp$ refers to the four spectral classes (see Table 1),
$L$ is the peak luminosity in the given energy band, 
$\Phi_0(L,sp)$ is the $z=0$ 
luminosity function of class $sp$, and $R_{\rm GRB}(z)$
the comoving GRB density distribution normalized at $z=0$. We 
assume that each \lfsp has a gaussian distribution of $\log L$
with a dispersion $\sigma_{\log L}$ around a central 
peak luminosity $L_c$. The GRB luminosity 
function \lf is the sum of the spectral luminosity functions \lfsp.

We assume that the photon spectrum is proportional to $E^{\alpha_{23}}$.
The peak flux $P(L,z)$ observed for a GRB of luminosity $L$ at redshift
$z$ is
\begin{equation}
P(L,z) = {L\over 4\pi A^2(z)}{(1+z)^{(2+\alpha_{23})}},
\end{equation}
where $A(z)$ is the bolometric luminosity distance for the cosmological
model. The peak flux \np distribution for GRBs of spectral class $sp$ is, 
\begin{equation}
N(>P,sp) = \int \Phi_o(L,sp)\,d\log L  
\int^{z(L,P,sp)}_0 R_{GRB}(z)(1+z)^{-1}\, (dV(z)/dz)\, dz\,
\end{equation}
where $z(L,P,sp)$ is derived from equation (2), $V(z)$ is the
comoving volume and the term $(1+z)^{-1}$ represents the time
dilation \citep{tot99}.
With the known distribution of flux limits $P_{lim}$ in the BD2 sample, 
we can derive the Euclidean value of \avvvm from the individual values
\vvm$ = (P/P_{lim})^{-3/2}$. The central peak luminosity $L_c$ of each
spectral class is iterated until \avvvm agrees with the observed value.

The limiting peak flux $P_{lim}$ depends on the GRB spectrum. Based on 
the BATSE detector response matrix, log $P_{lim}$ increases by 
$\sim 0.11$ from the softest to the hardest class. This does not
affect the \vvm derivation described above since $P$ is equally
affected. The luminosity $L_c$ derived for spectral class $sp$ scales
as $P_{lim}$.

The comoving GRB density distribution \sfr is often referred to as the 
'star formation rate' based on the expectation that GRBs are caused 
by massive stars. \citet{por00} have parametrized various models
for the evolution of the cosmic star formation rate (SFR) with
redshift. In model SF1, based on \citet{mad00}, the SFR rises rapidly
by an order of magnitude 
between $z=0$ and $z=1$, peaks between $z=1$ and $z=2$ and declines 
gently at higher redshifts. In model SF2, based on \citet{ste99}, it
rises similarly but then
remains roughly constant for $z>2$. Model SF3, reflecting the possibility
that extinction has been underestimated \citep{bla99}, has an SFR 
continuing to rise beyond $z=2$. For the cosmological model used in
this paper ($H_0 = 65~$km s$^{-1}$ Mpc$^{-1}$, $\Omega_M = 0.3$, 
and $\Omega_{\Lambda} = 0.7$), the star formation rates 
for $z = 1, 3, 5$ are 9.5, 7.7, 3.4 (SF1), 8.3, 12.8, 12.7 (SF2),
and 6.2, 12.8, 16.1 (SF3), respectively. We use these three models to
characterize \sfr in the derivation of the luminosity function.

\section{The GRB Luminosity Function}

\placefigure{fig3}

As descibed in the preceding section, we derive the luminosity function
at $z=0$ for each of the 4 spectral classes separately. We set the
gaussian dispersion of each \lfsp at $\sigma_{\log L} = 0.4$. 
The central peak luminosities $L_c$ of the spectral classes were 
determined from the corrected values of $<$\alp$>$ and \avvvm given 
in Table~\ref{tbl-1}. The resulting luminosity functions at $z=0$
for each of the four spectral classes 
for the SF2 model are shown in Figure~\ref{fig3}. 

\placefigure{fig4}

The sum of the spectral luminosity functions constitutes the overall 
luminosity function.  Figure~\ref{fig4} shows the resulting luminosity
functions for density distributions SF1, SF2, and SF3. The central peak 
luminosities $\log L_c$ range from $50.32-51.29$, $50.27-51.57$, and
$50.18-51.88$, respectively. The luminosity function generally appears 
to be a power law from $\sim 10^{50.5}$ erg s$^{-1}$ 
to $\sim 10^{51.5}$ erg s$^{-1}$, and then to decline 
more steeply. The total $z=0$ GRB densities are 0.48, 0.51, 
and 0.72 Gpc$^{-3}$ y$^{-1}$ for SF1, SF2, and SF3, respectively. 
The luminosities and densities quoted are 'isotropic-equivalent' values.
If all GRBs are beamed into, say, $\omega$ steradians, then 
luminosities require multiplication by $\omega/{4\pi}$ and
densities by $4\pi/\omega$. If the luminosity-hardness correlation
represents the distribution of luminosities within a GRB beam,
the situation would be more complex and the corrections to
luminosity and density would be a function of luminosity. 

\placefigure{fig5}

The cumulative distribution of peak fluxes observed in the BD2 sample 
is shown in Figure~\ref{fig5}. The predicted \np distributions are in 
excellent to good agreement with the observations. Compared to an
annual all-sky rate of 694 GRBs based on the BD2 sample, we expect 
above 0.1 (0.01) ph cm$^{-2}$ s$^{-1}$ annual rates of 2560 (5090), 
2720 (6810), and 2830 (8460) for cases SF1, SF2, and SF3, respectively. 

In Figure~\ref{fig6}, 
we show histograms of the expected redshift distribution in the BD2 sample. 
The fraction of high redshifts increases from SF1 to SF2 to SF3: the
expected fractions with $z>4$ are 1, 5, and 12\%, respectively. 
The largest single redshift that may be expected in the BD2 sample 
of 1391 GRBs on the basis of these three models is around 6, 13, or 
19, respectively. These, however, may be overestimates for SF2 and SF3,
since in these cases the star formation rate remains high at large
redshift, with no provision for the onset of star formation.

\placefigure{fig6}

In the derivation of the luminosity function, we assumed that 
the spectrum of the GRBs in each spectral classes was a simple
power law of slope $\alpha_{23}$. \citet{bel00} has compiled
GRB spectra as a function of a fluence hardness ratio, based on the
four BATSE LAD channels. Most of the spectra appear to be broken
power laws. The dependence on the hardness ratio seems to be a general
change in slopes with little change in break energy. We have
tested the effect of using these systematics by adopting instead
of the simple power law a Band spectrum \citep{ban93} with
$\alpha = \alpha_{23} + 1.0$, $\beta = \alpha_{23}$ and 
$E_0 = 150$ keV. The resulting luminosity function and predicted
distributions of flux and redshift were virtually identical to
that from the simple power law with slope $\alpha_{23}$.

\section{Luminosity-Hardness Correlation}

\placefigure{fig7}

Now that we have derived the GRB luminosity function, we are
in a position to generate the luminosity-hardness correlation
through a simulation. In order to reduce the effect of the
counting errors on \alp discussed in Sec. 2.1, we use the
\alp values of the brightest quartile, consisting of 348 GRBs 
with \avvvm$< 0.069$. We assigned each of these GRBs a random 
luminosity from its spectral luminosity function \lfsp. The
resulting plot of peak luminosity versus \alp is shown in
Figure~\ref{fig7}. For \alp$> -2.4$, we see clear evidence of 
the luminosity-hardness correlation, with a slope 
$d \log L/d \alpha_{23} \sim 1.1$. The dozen GRBs with
a large range of spectral indices below \alp$< -2.4$ belong 
to the softest spectral class and therefore based on the present
study cannot show any correlation between luminosity and hardness.

None of the GRBs observed by BATSE with redshifts listed by 
\citet{lam00} belong to the softest quartile in the BATSE catalog.
Therefore, we cannot at the present time check the main trend
from soft to hard bursts in Figure~\ref{fig7}. The total range in
luminosity of a factor of 50 in the \citet{lam00} list is compatible
with the range shown by the hard GRBs in Figure~\ref{fig7}. 
This may indicate {\it post facto} that the adopted value of 
$\sigma_{\log L} = 0.4$ is a reasonable one.

\section{Discussion}

\begin{figure}
\plotone{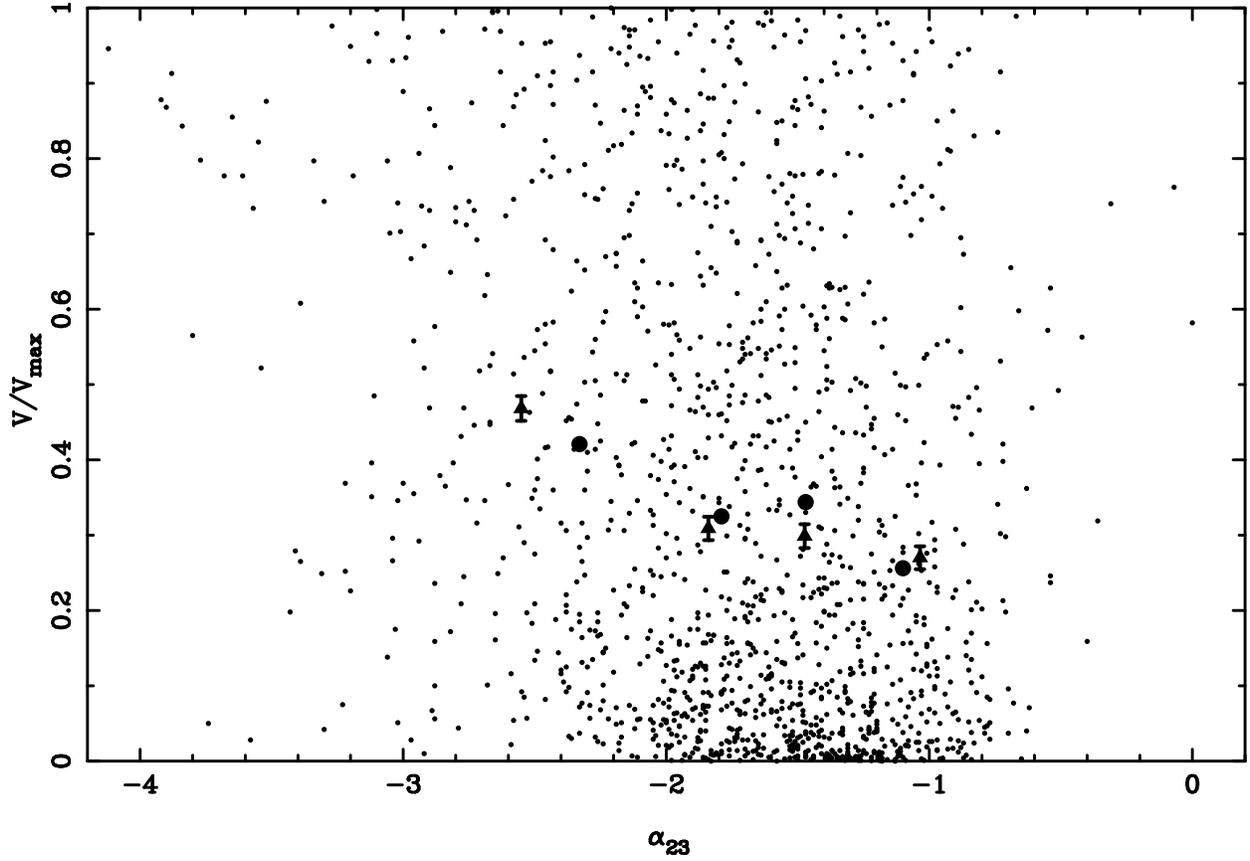}
\caption{Plot of Euclidean values of \vvm vs. spectral slope \alp for
1391 GRBs in the BD2 sample. The triangles show \avvvm for 4 spectral 
classes of $\sim348$ objects each with error bars denoting the mean 
errors of the \avvvm values. The large dots show the mean values
of $<$\alp$>$ and \avvvm corrected for the effect of statistical 
errors in the peak counts. \label{fig2}}
\end{figure}

\begin{figure}
\plotone{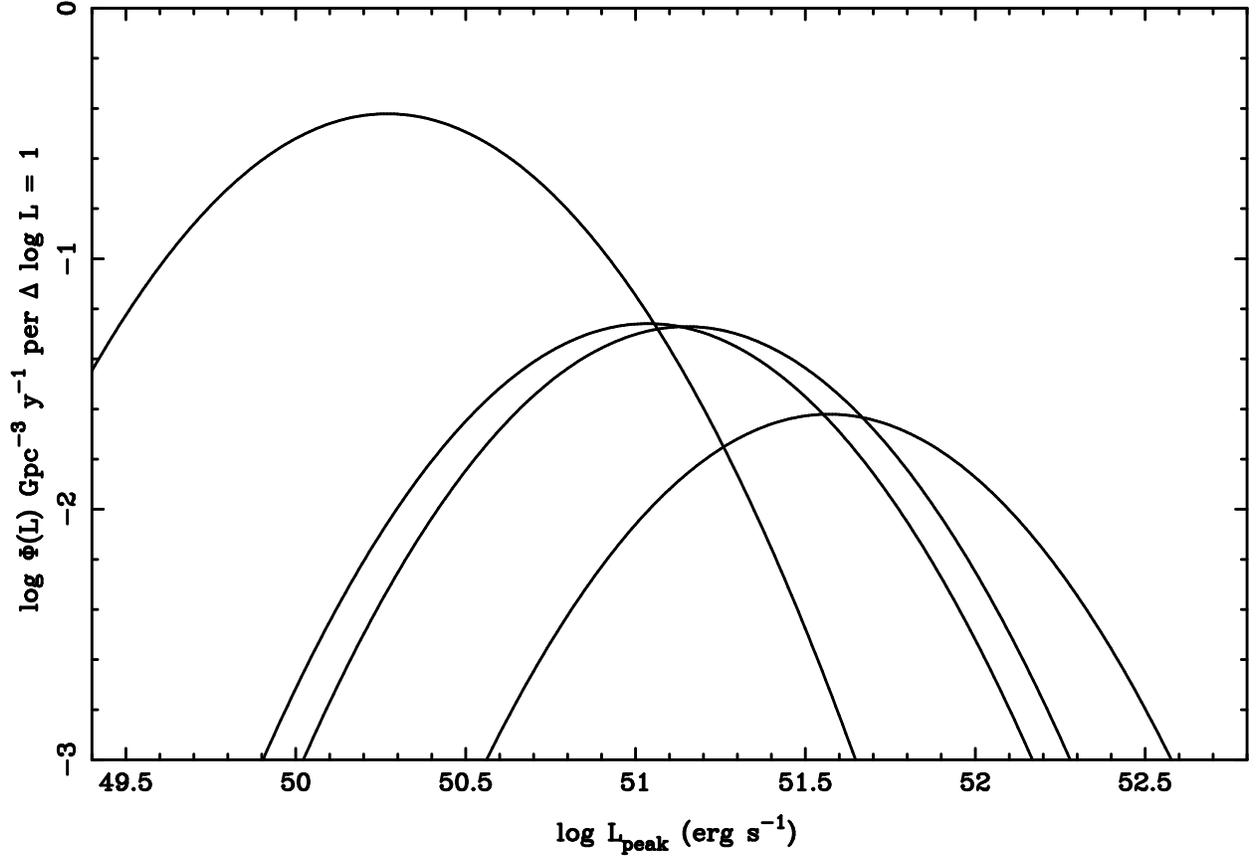}
\caption{Luminosity functions at $z=0$ derived for 4 
spectral classes for density distribution SF2.
$L_{\rm peak}$ is the peak luminosity in the energy range $50-300$ keV.  
Each of the luminosity functions is a gaussian with 
$\sigma_{\log L}$ = 0.4. The central luminosities are derived from the 
Euclidean \avvvm values. \label{fig3}}
\end{figure}

\begin{figure}
\plotone{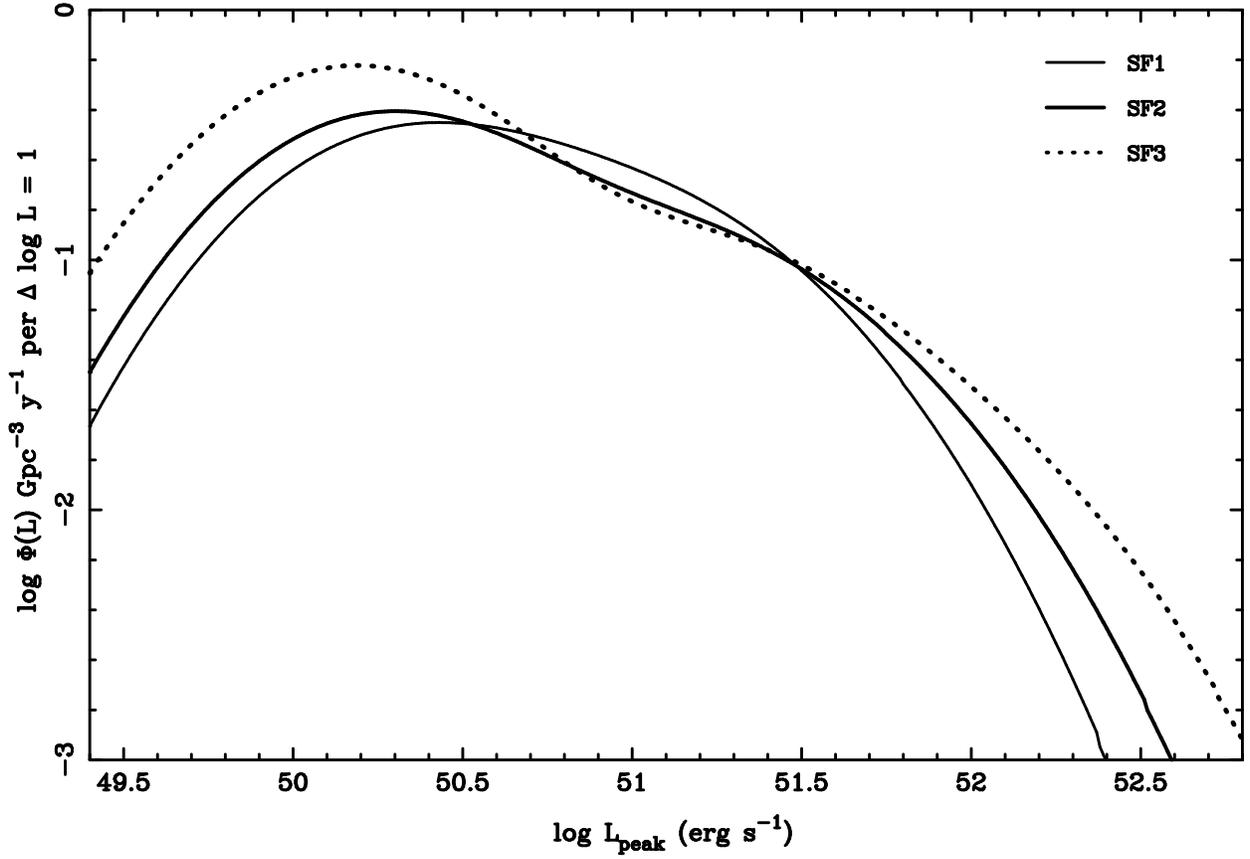}
\caption{GRB luminosity functions at $z=0$ for
density distributions SF1, SF2, and SF3, obtained as the sum of 
the 4 spectral luminosity functions. 
$L_{\rm peak}$ is the peak luminosity in the energy range $50-300$ keV.  
The luminosities and 
densities are 'isotropic-equivalent' values. \label{fig4}}
\end{figure}

\begin{figure}
\plotone{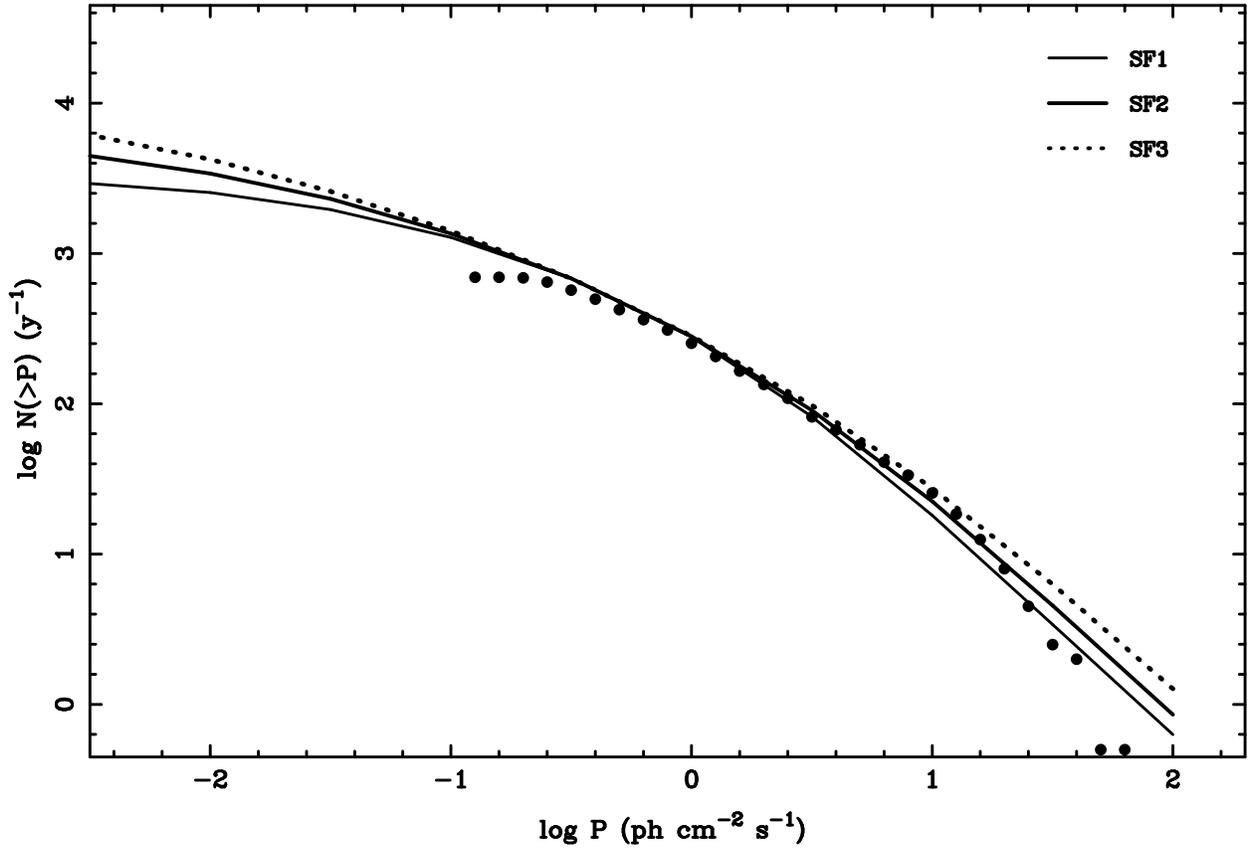}
\caption{Predicted \np distribution for GRBs 
based on the luminosity functions shown in Figure~\ref{fig4}. 
$P$ is the peak flux in the energy range $50-300$ keV.
The observed numbers in the BD2 sample are 
indicated as dots. \label{fig5}}
\end{figure}

\begin{figure}
\plotone{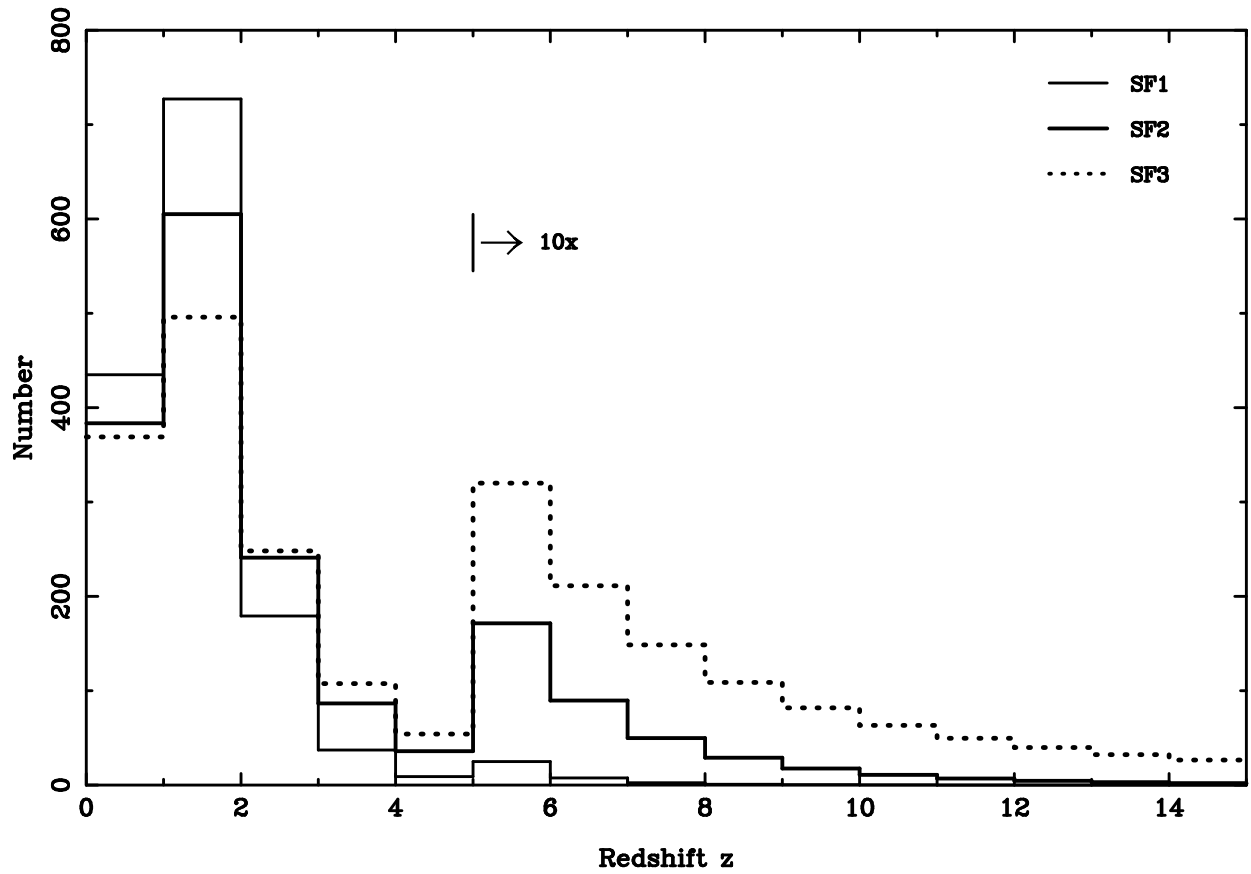}
\caption{Predicted redshift distribution for the 1391 
GRBs in the BD2 sample, based on the luminosity functions shown in 
Figure~\ref{fig4}. Numbers plotted are multiplied by 10 for $z>5$.
\label{fig6}}.
\end{figure}

\begin{figure}
\plotone{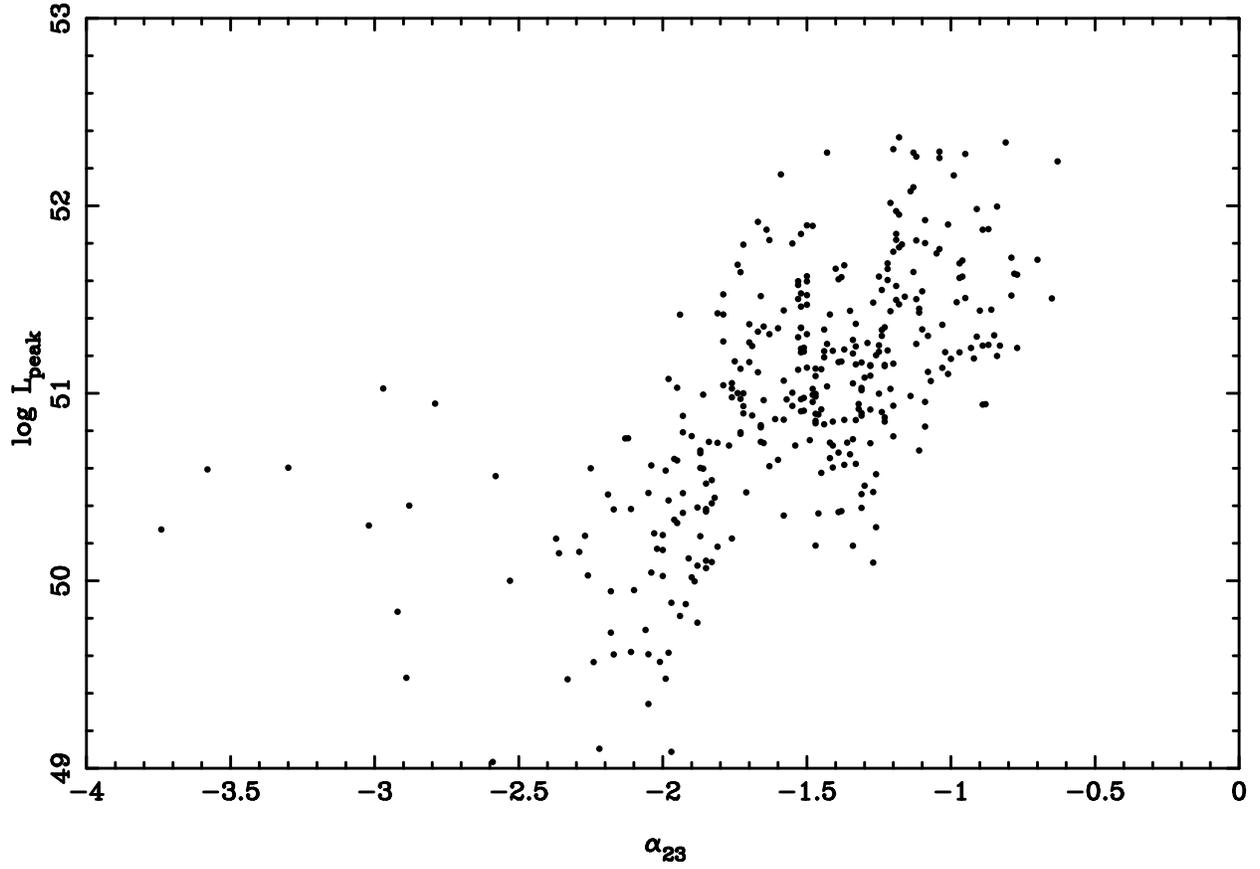}
\caption{Luminosity-hardness correlation of GRBs,
produced by a simulation using the luminosity function
based on the SF2 density distribution (see Figs. 3 and 4) 
and $\sigma_{\log L}$ = 0.4. \label{fig7}}
\end{figure}

We have shown that there exists for GRBs in the BD2 sample a 
$<V/V_{max}>$-hardness correlation, see Figure~\ref{fig2}.
Since the BD2 sample was selected from BATSE DISCLA data on
a timescale of 1024 ms, the correlation applies to GRBs with
durations larger than 1 s. Based on the realization that the 
Euclidean value of \avvvm is a cosmological distance indicator, 
we concluded that there exists a luminosity-hardness correlation, 
but could only demonstrate it after we had derived the luminosity 
function.

With the $<V/V_{max}>$-hardness correlation 
we were able to derive the luminosity function
without having to make assumptions about its overall shape,
as had been required till now \citep{sch99b}. We assumed a width
for the luminosity function of each spectral class that was
sufficient to produce a reasonably smooth overall luminosity function.
The entire excercise of deriving the luminosity function was
carried out without using any redshifts. We did have to make 
assumptions about the cosmological evolution or density 
distribution of GRBs for which we chose various star formation 
rate models.

This study has only been possible due to the power of the Euclidean 
value of \avvvm as an independent cosmological distance indicator.
Major progress beyond this work can be expected 
once systematic redshift surveys of GRBs become available. 
The luminosity-hardness correlation should become directly
observable. From samples with observed redshifts complete to 
given flux limits the GRB luminosity function can be derived 
without having to make assumptions about the density distribution. 
And while the Euclidean \avvvm cannot compete with the redshift 
as a distance indicator, the luminosity function should be 
compatible with the observed \avvvm values of the samples.

\acknowledgments

It is a pleasure to thank J. Brainerd, M. Finger and G. Pendleton
for information about the calibration of the BATSE detectors,
J. Tr\"umper for useful discussions and A. Iyudin for raising
questions about the statistical errors of the spectral index.

\clearpage

\begin{deluxetable}{ccccc}
\footnotesize
\tablecaption{\avvvm values as a function of $<\alpha_{23}>$ in the
BD2 sample.\label{tbl-1}}
\tablewidth{0pt}
\tablehead{
\colhead{number} & 
\colhead{$<\alpha_{23}>_{obs}$} & 
\colhead{\avvvm$_{obs}$} &  
\colhead{$<\alpha_{23}>_{corr}$} & 
\colhead{\avvvm$_{corr}$}   
}
\startdata
 348 & $-2.55$ & $0.468 \pm 0.017$ & $-2.33$ & $0.421$  \\
 348 & $-1.84$ & $0.309 \pm 0.016$ & $-1.79$ & $0.325$  \\
 347 & $-1.47$ & $0.299 \pm 0.016$ & $-1.47$ & $0.344$  \\ 
 348 & $-1.04$ & $0.270 \pm 0.015$ & $-1.10$ & $0.256$  \\

\enddata

\end{deluxetable}

\clearpage

\end{document}